\documentclass[12pt]{article}
\begin{document}

\title{Probable ratio of the vacuum energy in a Schwarzschild-de Sitter space}

\author{ Li Xiang\thanks{E-mail: xiang.lee@163.com}  and Y. G. Shen\\
 Shanghai Astronomical Observatory, Chinese Academy of Sciences,\\
 Shanghai, 200030, P. R. China\\
 and\\
 National Astronomical Observatories, Chinese Academy of Sciences,\\
  Beijing, 100012, P. R. China}

\date{}
\maketitle

\begin{abstract}
In this  Letter we extend the heuristic discussion of \cite{adler}
to a Schwarzschild-de Sitter spacetime ({\bf SdS}), based on the
generalized uncertainty principle({\bf GUP}). We try to explain
the present value of the vacuum energy density. An inequality is
put forward as an explanation for the smallness of the
cosmological constant, and the relationship between the Bekenstein
and the Bekenstein-Hawking entropy bounds is also briefly
discussed.



\end{abstract}
Although we do not possess a complete quantum theory of gravity,
some features which a complete theory is likely to possess may be
already embodied in well established laws of physics. For example,
taking into account gravity, Heisenberg's uncertainty principle
may be changed as \cite{adler}-\cite{cava}
\begin{eqnarray}\label{xp}
\Delta x\geq \frac{1}{\Delta p}+\lambda (\Delta p),
\end{eqnarray}
which implies a minimal length $2\sqrt{\lambda}$, where $\lambda$ is
of order of the Planck area $l_p^2\sim G$. The second term on
the right hand side of (\ref{xp}), the {\bf UV/IR} correspondence
\cite{cohen} term, was first derived from string theory \cite{veniq, gross},
from the scattering amplitude of a high energy string. Obviously,
this {\bf UV/IR} correspondence agrees with the features of a string,
since energy increases with its length. Some Gedankens
\cite{mag1}-\cite{adler1} verify this correspondence, and its algebra
has been investigated \cite{mag2,kempf}. Some effects of the generalized
uncertainty relation have been investigated by \cite{chang}, such
as deformation of the black body radiation spectrum,
and convergence of the vacuum energy (however, it is still
a factor $10^{120}$ greater than observational values). In
fact, \cite{rama} first pointed out the deformation of black body
radiation due to the GUP. However, \cite{rama} give an equation of state
different from \cite{chang} and the vacuum energy is still
divergent. The quantum correction to the entropy of a
black hole was investigated using this equation of state
in \cite{chang}, and the divergence appearing in the brick wall
model \cite{thooft} was found to be removed \cite{lix}.

The cosmological constant ($\Lambda$) problem is a challenge to
theoretical physics (see the recent review \cite{nahan2} and
references therein). As the vacuum energy density, the natural
value for $\Lambda$ will be of the same order as the Planck mass
squared, $\sim m_p^2$ \cite{hawkplb}. However, current astronomical
observations show that
\begin{eqnarray}
\frac{\Lambda}{m_p^2}\sim 10^{-120}.
\end{eqnarray}
We can treat $\Lambda$ as the ``effective vacuum energy" or
``effective cosmological constant" by introduction of a ``bare
cosmological constant" on the left hand side of Einstein's
equations. This is required to match the natural value of
$\Lambda$ with perfect accuracy, namely, ``fine tuning''. By what
mechanism is such fine tuning achieved? It has been argued that
the observed cosmological constant arises from quantum
fluctuations in the vacuum energy density rather than the vacuum
energy density itself \cite{nahan3}, however the mechanism of fine
tuning is still unclear. In Euclidean quantum gravity, the
transition amplitude between two field configurations on different
 hypersurfaces becomes divergent when $\Lambda\rightarrow 0$
\cite{hawkplb}. Hawking therefore argues that $\Lambda$ is
probably zero. The effects of a wormhole may also lead to a
vanishing of $\Lambda$ \cite{coleman, presk, hawknpb}. However,
evidence that the Universe is expanding at an increasing rate
implies that the cosmological constant is positive
\cite{carroll}\footnote{Amongst other candidates are quintessence,
as well as phantom and alternate theories of gravity, but
$\Lambda$ is the simplest choice.}.

If the {\bf GUP} is a fundamental element in quantum gravity, it should
qualitatively explain the smallness of the cosmological constant, at
least at a phenomenological level. The aims of this Letter are to
investigate the effects of the GUP on the thermodynamics of a
Schwarzchild-de Sitter spacetime, and discuss the implications
for the cosmological constant problem.

Let us start from the Schwarzschild-de Sitter geometry
\begin{eqnarray}
ds^2=-\left(1-\frac{2M}{r}-\frac{1}{3}\Lambda
r^2\right)dt^2+\left(1-\frac{2M}{r}-\frac{1}{3}\Lambda
r^2\right)^{-1}dr^2+r^2d\Omega,
\end{eqnarray}
where $\Lambda$ is the effective cosmological constant which
corresponds to the effective density of vacuum energy,
$\Lambda/8\pi$.\footnote{In the following discussions, when
talking about the vacuum energy or the cosmological constant, we
mean it is  ``effective".} The cosmological horizon $r_{c}$ is
given by
\begin{eqnarray}\label{horiz}
1-\frac{2M}{r_{c}}-\frac{1}{3}\Lambda r_{c}^2=0,
\end{eqnarray}
and thus the mass
\begin{eqnarray}\label{mass}
M=\frac{1}{2}\left(r_{c}-\frac{\Lambda}{3}r_{c}^3\right),
\end{eqnarray}
and the surface gravity at the horizon is given by
\begin{eqnarray}\label{kappa}
\kappa_{c}&=&-\frac{1}{2}\left(1-\frac{2M}{r}-\frac{1}{3}\Lambda
r^2\right)'_{r=r_{c}}\nonumber\\
&=&\frac{1}{2}\Lambda r_{c}-\frac{1}{2r_c},
\end{eqnarray}
where we have used Eq.(\ref{mass}). We can easily verify that
\begin{eqnarray}\label{firstlaw}
dM=-\frac{\kappa_c}{8\pi}dA_c-\frac{V}{8\pi}d\Lambda,
\end{eqnarray}
where $\Lambda$ is treated as a variable, and $V=4\pi r_c^3/3$ is
the volume of the Schwarzschild-de Sitter space\cite{cmp, huang}.
Note that Eqs.(\ref{horiz})-(\ref{firstlaw}) are valid almost to
the black hole, if $r_{c}$ is understood as the radius of the
hole's horizon, and only the definition of the surface gravity is
changed by a minus sign. For clarity, $r_c$ denotes the
cosmological horizon and $r_+$ the black hole horizon.

The thermodynamics of dS and asymptotic dS spacetimes have been
investigated in references \cite{gibbs}-\cite{bbwang}, where the
temperatures are still defined in terms of the surface gravity
of the horizons. It is difficult to define an equivalent temperature in
a SdS spacetime with a black hole. As Ref.\cite{cai2} points
out, we are in need of two conformal field theories to
describe the thermodynamics of a SdS spacetime with two horizons,
since the central charges of the Virasoro algebra are different for
the cosmological and black hole horizons. Here we follow
\cite{cvet}-\cite{bbwang} and introduce two temperatures
associated with the two horizons. A SdS spacetime is not
in thermodynamical equilibrium, but still static, and in order to
arrive at this, we set a heat-insulated spherical shell between
the cosmological and black hole horizons. Inside it, the black hole
is in equilibrium with a thermal bath of temperature
$\kappa_+/2\pi$. Outside it, the cosmological horizon is in
equilibrium with a thermal bath of temperature $\kappa_c/2\pi$.
Ref.\cite{bbwang} have investigated the thermodynamics of SdS
spacetimes by calculating the actions in two Euclidean sections.

Eq.(\ref{firstlaw}) cannot be regarded as the first law of
thermodynamics because of the problem of negative temperatures.
Furthermore, $M$ is not necessarily the mass of the black hole.
For instance, if there is an ordinary star without Hawking
radiation, $M$ is not a thermodynamic quantity related to
the Hawking radiation. Taking the total energy inside
the cosmological horizon $E_0=M+E_{vac}$ as conserved,
\begin{eqnarray}
0=dE_0=dM+dE_{vac},
\end{eqnarray}
which leads to the first law of thermodynamics for the
cosmological horizon
\begin{eqnarray}\label{evac}
dE_{vac}=-dM=\frac{\kappa_c}{8\pi} dA_c+\frac{V}{8\pi}d\Lambda.
\end{eqnarray}
The second term on the right hand side of the above equation is
attributed to the change in vacuum energy density, and represents
the variation of the ratio of vacuum energy to matter. Analogous
to chemical thermodynamics, it can also be attributed to the
chemical potential for the 4-volume of a Euclidean de Sitter
spacetime \cite{cmp,huang}. Thus, the inverse temperature and the
Bekenstein-Hawking entropy are given by, respectively,
\begin{eqnarray}
\beta&=&2\pi\kappa_{c}^{-1},\nonumber\\
S_{bh}&=&\frac{1}{4}A_c.
\end{eqnarray}
Similar formulae are still valid for a SdS black hole.

The preceding discussion does not involve corrections to the
temperature of the Hawking radiation due to the GUP. Strict
discussion on this requires a field theory dominated by the GUP.
Here we investigate this using an heuristic method first used by
Adler et al.\cite{adler}, who first considered the effects of the
GUP on the thermodynamics of a Schwarzschild black hole. Adler et
al. also suggest that the GUP may prevent the black hole
evaporating completely, resulting in a remnant with the Planck
mass. In \cite{adler}, the temperature of the black hole, i.e. the
characteristic energy of the Hawking radiation $\Delta E\sim
\Delta p$, is required to satisfy the position-momentum
uncertainty relation (\ref{xp}). If $\Delta x$ is identified as
the radius of the hole $r_h$, the modified temperature reads
\begin{eqnarray}\label{tempr1}
T\sim\Delta p\sim \frac{r_h-\sqrt{r_h^2-4\lambda}}{2\lambda},
\end{eqnarray}
which in the limit $\lambda\rightarrow 0$ gives the well known result
\begin{eqnarray}\label{tempr2}
T\sim \frac{1}{r_h}\sim \frac{1}{M}.
\end{eqnarray}
However, the argument of \cite{adler} is only valid for a
Schwarzschild black hole. For instance, the temperature of a
SdS hole is given by \cite{cai2}\cite{bbwang}
\begin{eqnarray}\label{tds}
T_{ds}=\frac{1}{4\pi}(\frac{1}{r_{+}}-\Lambda r_{+}),
\end{eqnarray}
where $r_{+}$ is the radius of the hole. Obviously, (\ref{tds}) is
different from (\ref{tempr1}) and (\ref{tempr2}). Let us look at
the generalized time-energy uncertainty \cite{gup}
\begin{eqnarray}\label{gupte}
\Delta t\sim \frac{1}{\Delta E}+\lambda (\Delta E),
\end{eqnarray}
which implies a minimum time element of order of the Planck scale
$(\Delta t)_{min}= 2\sqrt{\lambda}$. Solving (\ref{gupte}), we
have
\begin{eqnarray}\label{deltae}
\Delta E\sim \frac{\Delta t\pm \sqrt{(\Delta
t)^2-4\lambda}}{2\lambda}.
\end{eqnarray}
Since in the limit $\lambda\rightarrow 0$ the above equation should give
the usual time-energy uncertainty relation
\begin{eqnarray}\label{delt1}
\Delta E\sim \frac{1}{\Delta t},
\end{eqnarray}
we only take the minus--sign in (\ref{deltae}). The Hawking
temperature is proportional to the surface gravity $\kappa$,
which corresponds to an imaginary period $\beta=2\pi\kappa^{-1}$.
However, the temperature, i.e. the energy uncertainty of the hole
and the characteristic energy of particles emitting from the hole,
should satisfy (\ref{delt1}). Compared with $T_H=\beta^{-1}$, $\Delta t$
should be understood as the imaginary period of time $\beta$, in
a Euclidean section of the hole. Thus, taking into account the
GUP and substituting $\beta$ for $r_h$ in (\ref{tempr1}), we
obtain the modified temperature
\begin{eqnarray}\label{tempt1}
T&=&\frac{\beta-\sqrt{\beta^2-4\lambda}}{2\lambda}\nonumber\\
&=&\frac{2}{\beta+\sqrt{\beta^2-4\lambda}}.
\end{eqnarray}
We can also discuss the modified temperature in another way. It is
well known that Hawking radiation is a quantum effect, and that
the temperature is related to Planck's constant $\hbar$. In
normalised units where $G=c=1$, the Hawking temperature is given
by
\begin{eqnarray}\label{th}
T_H=\frac{\hbar\kappa}{2\pi}.
\end{eqnarray}
Recalling the usual uncertainty relation, $\Delta x\Delta p\geq
\hbar$, we define an ''effective" Planck constant,
$\hbar^{\prime}=\hbar[1+\lambda (\Delta p)^2]$. Eq.(\ref{xp})
therefore can be rewritten as $\Delta x\Delta p\geq
\hbar^{\prime}$. The effect of the GUP can then be taken into
account by substituting the ``effective" Planck constant for
$\hbar$,
\begin{eqnarray}
\hbar&\rightarrow&\hbar\left[1+\lambda(\Delta
p)^2\right]\nonumber\\
&\sim& \hbar\left[1+\lambda (\Delta E)^2\right].
\end{eqnarray}
Eq. (\ref{th}) is thus modified as
\begin{eqnarray}
T=\frac{\hbar\kappa\left[1+\lambda (\Delta E)^2\right]}{2\pi}.
\end{eqnarray}
We obtain Eq.(\ref{tempt1}) again, when the characteristic energy
$\Delta E$ as identified as the temperature $T$.

There is also an imaginary period, which corresponds to the
surface gravity of the cosmological horizon. Eq.(\ref{tempt1})
is also valid to the cosmological horizon if $\beta$ belongs to
the cosmological horizon. The entropy of the cosmological
horizon is given by
\begin{eqnarray}
S_c=\int T_c^{-1}dE_{vac},
\end{eqnarray}
where the integration is performed at constant $\Lambda$.

From (\ref{evac}), (\ref{tempt1}), and the relation
$\beta=2\pi\kappa_c^{-1}$, we obtain
\begin{eqnarray}\label{entryf}
S_c&=&\frac{1}{8}\int F[\kappa_c(r_c)]dA,\nonumber\\
F(\kappa_c)&=&1+\sqrt{1-\lambda\kappa_c^2/\pi^2}.
\end{eqnarray}
Similar formulae are still valid to the black hole horizon. The
total entropy of the SdS spacetime is the sum of the entropies of
the cosmological and black hole's horizon \cite{cai2,bbwang},
$S=S_c+S_+$. Considering the maximum entropy, we have $\delta
S=\delta S_c+\delta S_+=0$. On the other hand, energy conservation
demands
\begin{eqnarray}
0=T_c\delta S_{c}+T_{+}\delta
S_{+}+\frac{4\pi}{3}(r_c^3-r_{+}^3)\delta\Lambda,
\end{eqnarray}
then we have $T_c=T_+,~r_c=r_+$, namely,
\begin{eqnarray}\label{zero}
\kappa_c=\kappa_+=0,~ \Lambda=\frac{1}{r_c^2}=\frac{1}{r_+^2}.
\end{eqnarray}
Conditioned by (\ref{zero}), the entropies of the cosmological and
hole horizon approach their maximums. For example,
from (\ref{entryf}) we have
\begin{eqnarray}\label{maxi}
\delta S_c=0, \Leftrightarrow \frac{\partial F}{\partial
\kappa_c}=0, \Leftrightarrow \kappa_c=0,\nonumber\\
\Rightarrow
\frac{\partial^2
F}{\partial\kappa_c^2}=-\lambda/\pi^2,\Rightarrow\delta^2S<0.
\end{eqnarray}
Performing a similar procedure as from (\ref{entryf}) to
(\ref{maxi}), we can also show that the hole's entropy approaches
its maximum as $\kappa_+=0$. We notice that the vacuum energy is
given by
\begin{eqnarray}\label{evac2}
E_{vac}=\rho_{vac}V=\frac{\Lambda}{8\pi}\times \frac{4\pi
r_c^3}{3}=\frac{\Lambda r_c^3}{6},
\end{eqnarray}
then (\ref{mass}) can be rewritten as the following \cite{nahan1}
\begin{eqnarray}\label{rc2}
\frac{r_c}{2}=E_{vac}+M.
\end{eqnarray}
This is the total energy $E_0$ within the cosmological horizon,
including the energy of the matter and vacuum energy. Combining
(\ref{rc2}) and (\ref{evac2}) with (\ref{zero}), we obtain
\begin{eqnarray}
M=\frac{2}{3}E_0,~ E_{vac}=\frac{1}{3}E_0.
\end{eqnarray}
However, Eq.(\ref{zero}) is valid only for the existence of an
extreme black hole whose horizon coincides with the cosmological
horizon. This is an extreme black hole with zero temperature. It
means that the probable ratio of vacuum energy in a Schwarzschild-de
Sitter space is $1/3$, when the two horizons coincide. The
maximum of the total entropy is given by
\begin{eqnarray}
S_{max}=\frac{A}{2},
\end{eqnarray}
which violates the Bekenstein-Hawking entropy bound, but still
satisfies the D-bound $3\pi/\Lambda$ \cite{bousso}.

Although the maximum entropy is an unattainable limit, because the
third law of thermodynamics forbids a vanishing temperature, it
indicates the direction of the evolution of a Schwarzschild-de Sitter
spacetime. What is the situation when there is not a black hole
but an ordinary star? There is only a cosmological horizon, and
therefore from Eq.(\ref{entryf}) we obtain
\begin{eqnarray}\label{stdm}
S_c&=&\frac{1}{8}\int\left(1+\sqrt{1-\frac{4\lambda}{\beta^2}}\right)dA_c\nonumber\\
&\approx&\frac{1}{8}\int\left(2-\frac{2\lambda}{\beta^2}\right)dA_c\nonumber\\
&=&\frac{1}{4}A_c-\int\frac{\lambda}{4\beta^2}dA_c,
\end{eqnarray}
where $\beta^2>>4\lambda$. The first term is the
Bekenstein-Hawking entropy of the cosmological horizon, and the
second term
\begin{eqnarray}\label{correct}
\Delta S=-\frac{\lambda}{4}\int \frac{dA_c}{\beta^2}
\end{eqnarray}
is the correction due to the GUP. Substituting (\ref{kappa}) into
(\ref{correct}), we have
\begin{eqnarray}\label{ds}
\Delta S=-\frac{\lambda}{8\pi}(\ln r_{c}-\Lambda
r_{c}^{2}+\frac{\Lambda^2}{4}r_{c}^4).
\end{eqnarray}
The entropy is obviously related to the cosmological constant.
When fixing $r_c$, the entropy changes with $\Lambda$. In other
words, when the total energy is fixed, different ratios of
$\Lambda$ to matter make the entropies different. However, the
most probable ratio will make the departure from the maximum
entropy minimal, such that the minimal entropy production
principle be maintained. We therefore obtain
\begin{eqnarray}\label{smax}
S=\frac{A_c}{4}-\frac{\lambda}{16}\ln A_c+constant,
\end{eqnarray}
by setting
\begin{eqnarray}\label{partial}
\frac{\partial S}{\partial \Lambda}=0,\nonumber\\
\Lambda=\frac{2}{r_c^2}.
\end{eqnarray}
The logarithmic correction to the entropy is obtained, like the
black holes\cite{satoh}-\cite{carlip}. Substituting
(\ref{partial}) into (\ref{mass}), we have
\begin{eqnarray}\label{matt}
M=\frac{r_c}{2}\times \frac{1}{3},\nonumber\\
E_{vac}=\frac{2}{3}E_0,
\end{eqnarray}
which roughly agrees with results derived from observations,
($\Omega_{\Lambda}, \Omega_{m}$)=(0.7,0.3) \cite{carroll}.

The above discussions ignore the entropy of matter, since this
part is very much less than the contribution from the cosmological
horizon. Furthermore, it is reasonable to suppose that the entropy
of matter is independent of the vacuum energy. Thus, the final result
will not change, even if the entropy of the matter is taken into
account. Effectively, the matter is assumed to be in a pure state
with vanishing entropy.

What else can the {\bf GUP} tell us?

In a regime dominated by the {\bf UV/IR} duality, i.e. of ultra
high energy, we have
\begin{eqnarray}
\Delta t >\lambda \Delta E,
\end{eqnarray}
and the luminosity (or the energy loss rate) of a system reads
\begin{eqnarray}\label{rate1}
\frac{dE}{dt}\sim \frac{\Delta E}{\Delta t}< \frac{1}{\lambda}\sim
G^{-1}.
\end{eqnarray}
On the other hand, the energy loss rate
\begin{eqnarray}\label{rate2}
\frac{dE}{dt}=\frac{1}{4}\rho A\sim \rho\pi R^2,
\end{eqnarray}
where $\rho$ is the energy density of the system. The variable $A$ is
the area and $R$ the length of the system. From
(\ref{rate1}) and (\ref{rate2}), we obtain an inequality which
associates the energy density with the length of the system
\begin{eqnarray}\label{rho1}
\rho<\frac{1}{\lambda R^2}.
\end{eqnarray}
It is compatible with the effective field theory with the duality
of the UV and IR cutoffs $\varepsilon$ and $L$,
$L^2\varepsilon^4\leq m_p^2$ \cite{cohen}. Since $\varepsilon^4$
is the maximum energy density, Eq.(\ref{rho1}) can be regarded as
evidence for Ref.\cite{cohen}. Furthermore, from (\ref{rho1}) we
obtain a small vacuum energy density, $\rho_{vac}\sim H_0^2$, if
$R$ is regarded as the present size of the Universe, $ R\sim
t_0\sim H_0^{-1}$, where $t_0$ is the age of the Universe and
$H_0$ is the present Hubble parameter.

The holographic principle \cite{hooft, susskind}, relating the dual
field theory on the boundary and gravity in the bulk, is
regarded as an elementary principle. An important feature of it is
that the entropy of a system of matter cannot exceed the
Bekenstein-Hawking entropy bound. (Recent progress has been
discussed in \cite{bousso}, and some other entropy bounds have been
proposed). We believe that the GUP is closely related to the
holographic principle, because: (a) the GUP modifies the density
of quantum states, and the entropy divergence arising from
the brick wall model is removed; (b) Ref.\cite{cohen}
argues that the Bekenstein-Hawking entropy bound requires an
effective field theory with UV/IR correspondence compatible with
the GUP. Here we discuss the relation between the Bekenstein bound
(without gravity) and the Bekenstein-Hawking bound.

When Bekenstein proposed his famous entropy bound \cite{beken}
\begin{eqnarray}
S\leq 2\pi ER,
\end{eqnarray}
he did not take into account gravity. We need an additional
relationship to associate the Bekenstein bound with
the Bekenstein-Hawking bound
\begin{eqnarray}
S\leq \frac{A}{4G},
\end{eqnarray}
where $G$ is Newton's constant. Eq.(\ref{rho1}) is such a
relationship, which leads to
\begin{eqnarray}\label{rho2}
E<\lambda^{-1}R.
\end{eqnarray}
Substituting it into Bekenstein's bound, we have
\begin{eqnarray}
S\leq \frac{\pi R^2}{\lambda}\sim \frac{A}{4G}.
\end{eqnarray}

In summary, We have made some heuristic discussions about the
thermodynamics of SdS spacetimes and the cosmological constant
problem, starting from the generalized uncertainty relation
(\ref{xp}). Based on the work of \cite{adler}, we have derived
corrections to the temperature and entropy of the cosmological
horizon using an heuristic method. The logarithmic correction to
the Bekenstein-Hawking entropy was also obtained. For an usual
star, the probable ratio of vacuum energy to the total energy is
$2/3$. The generalized uncertainty principle leads to a bound on
the energy loss rate for physical systems. A relation between the
energy density and length was suggested and utilized to explain
the small cosmological constant, and associate the Bekenstein
entropy bound with the Bekenstein-Hawking entropy.

Although a SdS spacetime does not describe the real Universe, the
qualitative results derived here are also believed to be valid
for a FRW geometry. Consider the following Gedanken: the matter of
a usual star is distributed in an homogeneous and isotropic FRW
universe with positive $\Lambda$, by some reversible and adiabatic
processes. According to thermodynamics, these processes are isentropic.
Furthermore, such classical isentropic processes should not
change the ratio of vacuum energy. This is because a decaying
$\Lambda$ leads to an increase in entropy, which contradicts
the nature of an adiabatic and isentropic process. Therefore, the
probable ratio of vacuum energy in a SdS spacetime is also
valid in a FRW universe. We prefer the thermodynamics of a SdS
spacetime without a black hole. After all, the Universe is not
dominated by black holes. Although the mass of a black hole
can be distributed in a FRW universe by the Hawking effect,
some thermodynamical properties of a black hole, such as the
heat capacity and the super additivity of the entropy, are essentially
different from thermal radiation and usual matter. Furthermore, we do
not know whether a reversible and adiabatic process is well defined
for the transition from black hole to radiation.

However, Eq.(\ref{matt}) corresponds only to the greatest
probability. We cannot exclude the possibility that the
cosmological constant has other values. We have noticed that
Eq.(\ref{matt}) is not completely consistent with observations. In
fact, the weight of the vacuum energy (or dark energy) is
constrained by the current age of the Universe\cite{nahan2}. But
derivation of (\ref{matt}) is independent of the age, thus it is
not a conclusive result. Other uncertain factors could possibly
affect the weight of the vacuum during the evolution of the
Universe.

We note that the formation of a black hole tends to decrease the
weight of the vacuum energy. This is contrary to intuition: the
black hole should shrink because its temperature is greater than
the temperature of the cosmological horizon. However, Bousso \&
Hawking observed \cite{bousso2} and Nojiri \& Odintsov confirmed
\cite{odintsov} that there are two modes of perturbation,
corresponding to the evaporation and {\bf anti-evaporation} of a
SdS black hole. In the latter, a drastic effect, the energy from
the cosmological horizon flows into the black hole, and may result
in a decrease in the ratio of vacuum energy. However, according to
our understanding, anti-evaporation occurs only in a system near
equilibrium. As we have discussed from (\ref{entryf}) to
(\ref{maxi}), the second law of thermodynamics prefers the
anti-evaporation mode of a nearly extreme black hole whose size is
comparable to the cosmological horizon. The black hole whose size
is much smaller than the cosmological horizon, is still dominated
by evaporation, because it is far away from thermodynamic
equilibrium.

Another possibility is that the properties of the background are
influenced by the black hole: the black hole leaves its impression
on $\Lambda$ and shows a sign of nonlocality. Why so? We
hypothesize that it is possibly associated with the black hole
information puzzle \cite{presk2}: can the unitarity of quantum
theory be maintained during the evaporation of a black hole? The
solution to this puzzle requires a nonlocal field theory (such as
string theory) \cite{lowe}. Recently, it has been argued that the
thermal radiation and entropy of a black hole arise due to
decoherence. Namely, there exist correlations between the hole and
its environment \cite{inf2,inf3}. If so, the environment
surrounding the black hole, the de Sitter space, should be
influenced  by the black hole formation. Decoherence is based on
the quantum entanglement of a system with its environment. It
cannot be identified with, and must not be confused with
dissipation \cite{joos}. However, decoherence often occurs
together with dissipative phenomena, such as particle creation
from a gravitational field \cite{cast}. If the formation of a
black hole induces cosmological constant decay and then produces
particles, the information-loss puzzle is avoided by the
correlation between these particles and the black hole. This is
because a pure state remains pure only for a closed system.
Certainly, it is only a hypothesis, and the mechanism of
cosmological constant decay is still unclear. While it is
desirable that these issues are investigated further in a rigorous
manner, we leave them for future research.

\section*{Acknowledgments}
The authors would like to thank Peter Williams who read our
manuscript and improved the English.
 This research is supported by NSF of China (Grants No. 10273017 and No. 10373003) and the Foundation of Shanghai Development for Science and Technology (Grant No.01-JC14035).

\end{document}